# Musical NeuroPicks :

# a consumer-grade BCI for on-demand music streaming services

**Fotis P. Kalaganis[1], Dimitrios A. Adamos[2,3], Nikos A. Laskaris[1,3]**

[1]AIIA Lab, Department of Informatics,

[2]School of Music Studies

[3]Neuroinformatics GRoup, http://neuroinformatics.gr

**Aristotle University of Thessaloniki**

54124 Thessaloniki, Greece

E-mail: kalaganis@csd.auth.gr; dadam@mus.auth.gr; laskaris@aiia.csd.auth.gr

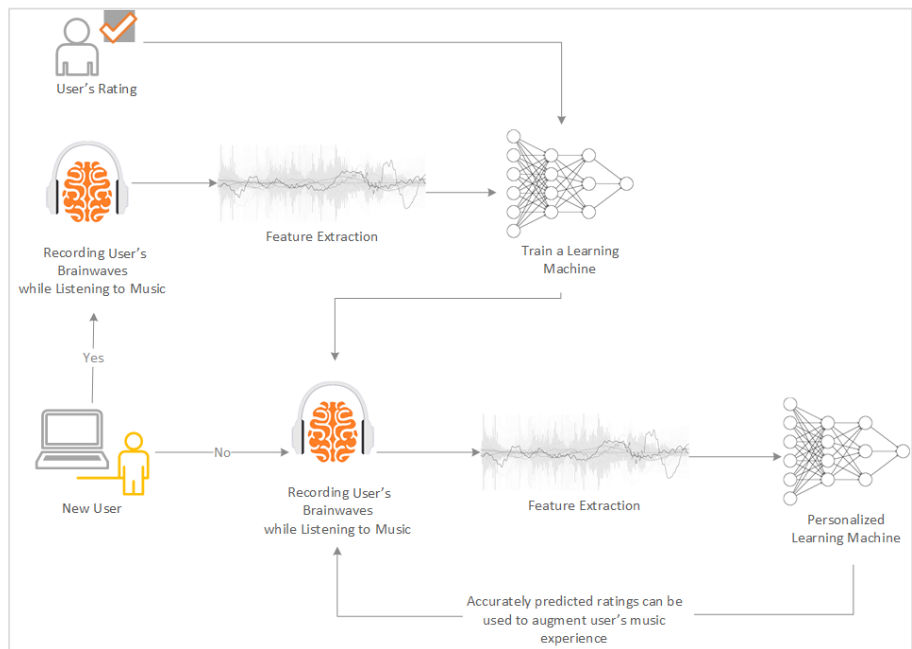

## Abstract

We investigated the possibility of using a machine-learning scheme in conjunction with commercial wearable EEG-devices for translating listener's subjective experience of music into scores that can be used in popular on-demand music streaming services.

Our study resulted into two variants, differing in terms of performance and execution time, and hence, subserving distinct applications in online streaming music platforms. The first method, NeuroPicks, is extremely accurate but slower. It is based on the well-established neuroscientific concepts of brainwave frequency bands, activation asymmetry index and cross frequency coupling (CFC). The second method, NeuroPicks$^{VQ}$, offers prompt predictions of lower credibility and relies on a custom-built version of vector quantization procedure that facilitates a novel parameterization of the music-modulated brainwaves.

Beyond the feature engineering step, both methods exploit the inherent efficiency of extreme learning machines (ELMs) so as to translate, in a personalized fashion, the derived patterns into a listener's score. NeuroPicks method may find applications as an integral part of contemporary music recommendation systems, while NeuroPicks$^{VQ}$ can control the selection of music tracks. Encouraging experimental results, from a pragmatic use of the systems, are presented.

# 1 Introduction

Until recently, electroencephalography (EEG) was met exclusively in hospitals and clinics, where trained experts operated expensive devices. The vast majority of the EEG-related research was dedicated to the diagnosis of epilepsy, sleep disorders, Alzheimer's disease, as well as the monitoring of certain clinical procedures such as anesthesia. By the same token, the application of Brain-Computer Interfaces (BCIs) has so far been confined to neuroprosthetics and for building communication channels for the physically impaired people [1].

Recent advances in biosensors [2] offer commercial neuroimaging headsets at affordable prices. Nowadays, the procedure of recording the electrical activity of the brain via electrodes on the human scalp has been significantly simplified and does not require a clinician's level of expertise [3]. This emerging potential stimulates growth, favors innovation and anticipates novel applications of non-invasive BCIs within real-life environments [4].

Since the beginning of the twenty-first century, the digital revolution has radically affected the music industry and is continuously reforming the business model of music economy [5]. So far, well-established channels of music distribution have been replaced and new industry stakeholders have emerged. Among them, music on-demand recommendation and streaming services emerge as the "disruptive innovators" [6] of the new digital music ecosystem.

In a previous work [7], we have introduced our vision for the integration of bio-personalized features of musical aesthetic appreciation into modern music recommendation systems to enhance user's feedback and rating processes. A feasibility study was then carried out, based on the recording of brain activity via a 14-electrode modern commercial wireless EEG device. Therein, a single-sensor discernible EEG pattern that reflected the personalized musical aesthetic experience was identified over the prefrontal cortex and a proof of concept was provided by means of regression analysis. In the current work, we aimed for a more systematic development of a machine learning scheme that could harness the information from the brain activity recorded without sacrificing the convenience of the listener. The BCI system we opted for is presented pictorially in Fig. 1. Among the desired specifications was the seamless incorporation of a wearable EEG device within an online streaming music service.

We have now employed an affordable 4 dry-sensor wireless EEG device that operates with higher resolution and focuses on the prefrontal cortex.

The first part of our study was devoted to the identification of the optimal brain activity descriptors that, within the range of capabilities provided by the device, would reliably and consistently reflect the listeners' evaluation about the music being played. We experimented with a set of standard signal-descriptors, conventionally employed in human electroencephalography, that could readily fit in a real-time system. They were based on single-sensor measurement and satisfied the requirements of low computational cost. A particular combination of descriptors was detected that offered very high accuracy and selected as a composite biomarker of music appreciation. Since its performance was reaching satisfactory levels -only- when applied to long temporal segments of brain activity, we resorted to an alternative subject-adaptive descriptor that could facilitate a music-evaluation BCI-system operating at a faster pace. Tailored to the task of music evaluation, appears as a powerful novel biomarker consistent with the nonstationarities and nonlinearities of brain activity.

The second part of this study concerned the interface of either biomarker with an extreme learning machine and the implementation/validation of the overall music-evaluation BCI system. Two distinct scenarios appeared as covered by the proposed approach. The first one relates with the need in modern music streaming services for accurate ratings by the user that will, in turn, be exploited in order to deliver suggestions that match the user's taste. The second one, relates with the necessity for an effortless interaction of the users with the application's interface, for instance, so as to change the currently delivered song whenever this is not among their preferences. For this case the BCI-system should be able to provide a prompt estimate (i.e. a prediction of the final music evaluation score) at a tolerable error. An ELM in either case takes over the translation of brain activity descriptors into a single numbered score expressing the appraisal of music within the range 1-5. The two introduced variants of BCI-system, namely NeuroPicks and NeuroPicks$^{VQ}$, can be personalized with very limited amount of training and their operation induces negligible amount of delay.

In our experimentations, we adopted a passive listening paradigm utilizing the *Spotify* on-demand music streaming service (http://spotify.com). Overall, the outcomes of this work are very encouraging for conducting experiments about music perception in real-life situations and embedding brain signal analysis within the contemporary technological universe.

A preliminary version of this work, describing only the NeuroPicks system, has been presented in [8]. The need for an additional, supposedly complementary biomarker that could reveal the users' intentions, was the main motivation for this extended version. The NeuroPicks system takes advantage of well-established features that describe human cognition processes, while the NeuroPicks$^{VQ}$ system exploits a novel descriptor of nonlinear brain response dynamics EEG.

The remaining paper is structured as follows. Section 2 serves as an introduction to EEG and its role in describing and understanding the effects of music. Section 3 outlines the essential tools that were employed during data analysis. Section 4 describes the experimental setup and the adopted methodology for analyzing EEG data. Section 5 is devoted to the presentation of results, while the last section includes a short discussion about the limitations of this study and its future perspectives.

## 2 Electroencephalography and Music perception studies

Electroencephalography is a non-invasive neuroimaging technique that monitors the electrical activity of the brain using electrodes placed on the scalp. EEG reflects mainly the summation of excitatory and inhibitory postsynaptic potentials at the dendrites of ensembles of neurons with parallel geometric orientation. While the electrical field produced by distinct neurons is too weak to be recorded with surface EEG electrode, as neural action gets to be synchronous crosswise a huge number of neurons, the electrical fields created by individual neurons aggregate, resulting to effects measurable outside the skull [9].

The EEG brain signals, also known as brainwaves, are traditionally decomposed (by means of band-pass filtering or a suitable transform) and examined within particular frequency bands, which are denoted via Greek letters and in order of increasing central frequency are defined as follows: δ (0.5-4)Hz, θ (4-8)Hz, α (8-13)Hz, β (13-30)Hz, γ (>30)Hz. EEG is widely recognized as an invaluable neuroimaging technique with high temporal resolution. Considering the dynamic nature of music, EEG appears as the ideal technique to study the interaction of music, as a continuously delivered auditory stimulus, with the brainwaves. For more than two decades neuroscientists study the relationship between listening to music and brain activity from the perspective of induced emotions [10, 11, 12]. More recently, a few studies appeared which shared the goal of uncovering patterns, lurked in brainwaves, that correspond to subjective aesthetic pleasure caused by music [13, 14].

Regarding music perception, the literature has reported a wide spectrum of changes in the ongoing brain activity. This includes a significant increase of power in β-band over posterior brain regions [15]. An increase in γ band, which was confined to subjects with musical training [16], an asymmetrical activation pattern reflecting induced emotions [17] and an increase of frontal midline θ power when contrasting pleasant with unpleasant musical sounds [18].

Regarding the task of decoding the subjective evaluation of music from the recorded brain activity, the role of higher-frequency brainwaves has been identified [19], and in particular the importance of γ-band brainwaves recorded over forebrain has been reported [20]. More recently, our group has exploited the concept of nested oscillations in the brain [7] to introduce a relevant CFC biomarker for the assessment of spontaneous aesthetic brain responses during music listening. The reported experimental results indicated that β and γ EEG oscillations recorded over the left prefrontal cortex are crucial for estimating the subjective aesthetic appreciation of a piece of music and may reflect the interconnectivity of the frontal cortex with subcortical music-rewarding dopaminergic areas.

## 3 Methods

This section describes the methodological elements employed towards the design and realization of the introduced BCI-systems. It commences by presenting, briefly, the descriptors of neural activation that were examined on their ability to convey the aesthetic appreciation during music listening. These descriptors were chosen among an extended repertoire which is currently in use in neuroscientific studies. Additionally, a novel descriptor of brain dynamics is introduced, based on the reparameterization of brainwaves according to a learned dictionary of short-term activations. The potential of each individual descriptor (and their combinations) to mediate the desirable read-out was estimated using *Distance Correlation*, which is described next. Finally, the selected machine learning scheme, that incorporated an important class of artificial neural network (ANNs), is described.

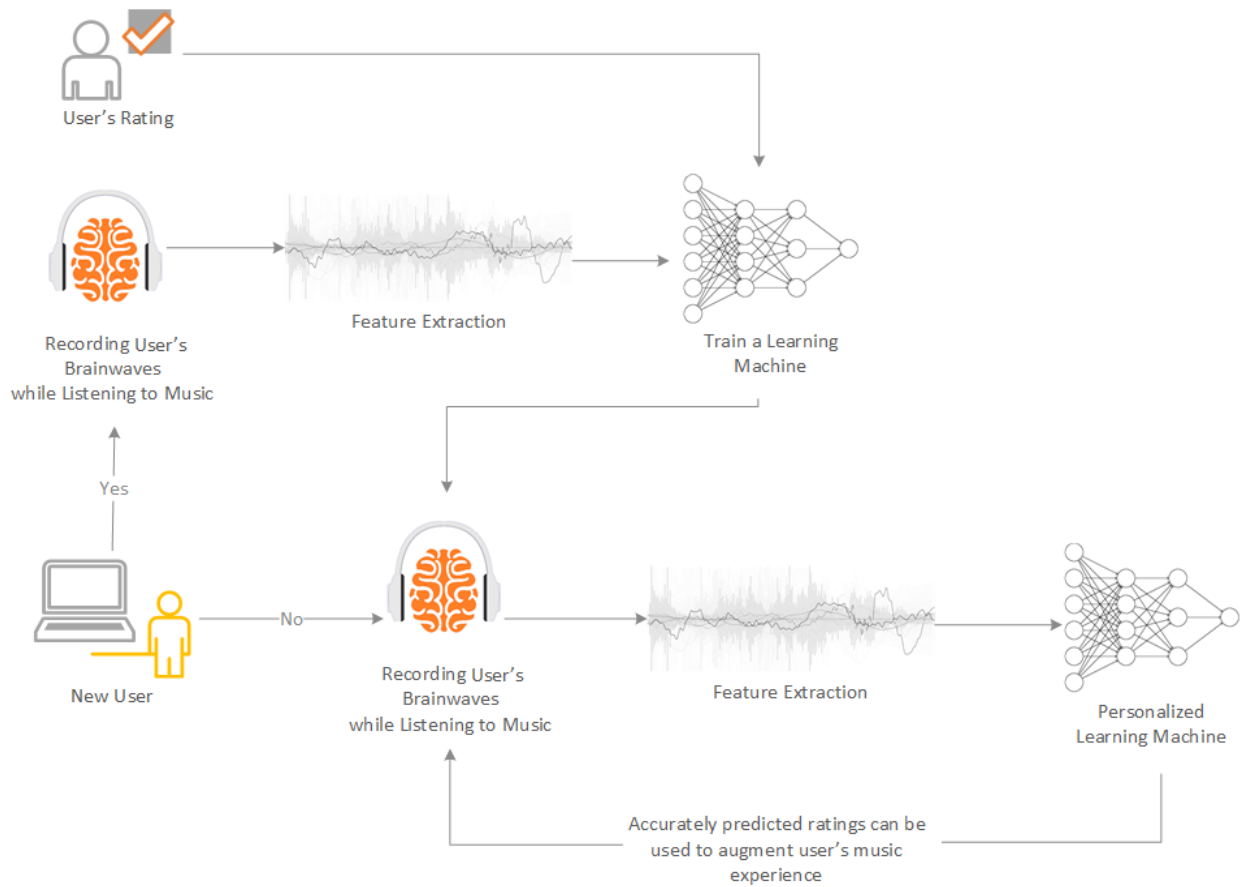

*Figure 1* Flow chart of the desirable music evaluation BCI

3.1 Neural activation profiling -The composite biomarker

Brainwaves are often characterized by their prominent frequency and their (signal) energy content. Here, we adopted a quasi-instantaneous parameterization of brainwaves content, by means of Hilbert transform. The signal from each sensor $x(t)$, was first filtered within the range corresponding to the frequency-band of a brain rhythm (like δ-rhythm) and the envelope of the filtered activity $A_{rhythm}(t)$ was considered as representing the momentary strength of the associated oscillatory activity. Apart from the amplitude of each brain rhythm, its relative contribution was also derived by normalizing with the total signal strength (summed from all brain rhythms).

In neuroscience research, activation refers to the change in EEG activity in response to a stimulus and is of great interest to investigate differences in the way the two hemispheres are activated [21]. To this end, an activation asymmetry index was formed by combining measurements of activation strength from two symmetrically located sensors, i.e. $AI(t) = {}^{left}A_{rhythm}(t) - {}^{right}A_{rhythm}(t)$. The normalized version of this index was also employed as an additional alternative descriptor.

A third descriptor was based on the CFC concept, which refers to the functional interactions between distinct brain rhythms [22]. A particular estimator was employed [23] that quantified the dependence of amplitude variations of a high-frequency brain rhythm on the instantaneous phase of a lower-frequency rhythm (a phenomenon known as phase-amplitude coupling (PAC)). This estimator operated on each sensor separately and used to investigate all the possible PAC couplings among the defined brain rhythms.

It is important to notice here, that the included descriptors were selected so as to cover different neural mechanisms and share a common algorithmic framework. Their implementation -and mainly their integration- within a unifying system did not induce time delays unreasonable for the purposes of our real-time application since they are based on band-pass filtering and Hilbert Transform for deriving the instantaneous amplitude and phase.

### 3.2 Nonlinear Dynamics Descriptor – The new biomarker

In an attempt to reveal the subjective music preference, we introduce an additional descriptor, which is based on a semi-supervised data-learning scheme that adapts to the music-modulated brain dynamics of the user. The descriptor stems from the blending of nonlinear dynamics principles with pattern analysis concepts, and experimentally proved to outperform the well-established descriptors when operates on short signal segments. The overall approach builds over the idea of reconstructing brain dynamics from the recorded sensor signal, representing them as trajectories in a high-D space and provide a computable relevant parametrization, by means of data learning, that incorporates the listener's appraisal scores. This representation follows a discriminative VQ scheme, which was introduced in [24] for disentangling pathological from healthy cognitive responses. The critical modification, that is incorporated here, is the codebook editing step which is based on correlation analysis of the encoded information with the provided scores.

### 3.2.1 VQ-encoding of single-trial traces

The first step of reconstructing dynamics, is performed by means of time-delay embedding. The pair of parameters of time-delay and embedding dimension ($\tau$, $d_e$) is set to appropriately selected values, i.e. a small $\tau$ to preserve the temporal resolution of the signal and sufficiently large $d_e$ so as to track the intrinsic dynamics. Then a given 1D signal $x(t)$, $t=1,\ldots,T$ is represented as a trajectory in a phase-space by forming the consecutive vectors $\mathbf{x_i}=[x(t_i), x(t_i+\tau), x(t_i+2\tau), \ldots, x(t_i+d_e\tau)]$, $i=1,\ldots,T-d_e\tau$. The trail of this trajectory is a set of $N = T - d_e\tau$, time-indexed, $d_e$-dimensional points and equivalently represented in the tabular form of a trajectory matrix: **Tr**{ $x(t);d_e,\tau$ }=[$\mathbf{x_1}$| $\mathbf{x_2}$| ... | $\mathbf{x_N}$] , where ("|" denotes the row separator). The formed trajectory matrix encapsulates the brain response dynamics in "raw" format. A more refined signature can be derived by means of vector-quantization (VQ). Given a set of reference points in phase-space, a partition -known as Voronoi tessellation- is induced that enables the description of the trajectory-trail in the form of a distribution pattern. Fig.2 exemplifies these steps based on 3 different single-trial signals recorded from a subject during music listening. The original signals, after band-pass filtering (Fig.2a), have been represented as 2D trajectories in Fig.2b. In the same figure, 10 reference points are also depicted together with the corresponding Voronoi diagram. The shown Voronoi regions can be thought of as 2D bins and have been used in the re-parameterization of each trajectory in the form of a histogram (see Fig.2c).

Provided a codebook (i.e. a set of reference vectors, also known as codevectors) in reconstructed phase-space, the algorithmic procedure of **VQ-encoding**, involves the application of the 'nearest-neighbor rule' to each row-vector in the trajectory matrix so as to assign it to the most similar codevector. Formally, the codebook $\{c_j\}_{j=1,\ldots,k}$ divides the phase-space into k Voronoi regions, $V_i \subset \mathbb{R}^{d_e}$, i=1,...,k

$$V_i = \left\{ \mathbf{x} \in \mathbb{R}^{d_e} : \| \mathbf{x} - \mathbf{c_i} \|_2 \leq \| \mathbf{x} - \mathbf{c_j} \|_2, \forall j, j = 1, \ldots, i-1, i+1, \ldots, k \right\}$$

and the single-trial encoding step transforms a reconstructed trajectory to a distribution pattern over these regions. It is prominent, that the design of codebook is a critical part towards the definition of a suitable descriptor that would emphasize the characteristics of brain dynamics that reflect the level of music appraisal.

### 3.2.2 Codebook Design and selection of CodeWaves

The codebook design is based on a data-learning algorithmic scheme and presumes the availability of a certain amount of training data, i.e. subject-specific EEG recordings during music listening along with the subjective evaluation score. The procedure begins with the execution of Neural Gas algorithm for the initial design of codebook and proceeds with the selection of the most informative codevectors. The first stage, that is the derivation of prototypical brainwave patterns, runs in unsupervised mode and captures the underlying dynamic manifold by demarcating (a user defined number of) k Voronoi regions. The second stage, that is the selection of regions reflecting better the music appraisal, is based on correlating the ''activation'' of each Voronoi region with the subjective score. The overall scheme (see Fig.3a) is of a semi-supervised nature and enhances further the personalization of the proposed BCI-system.

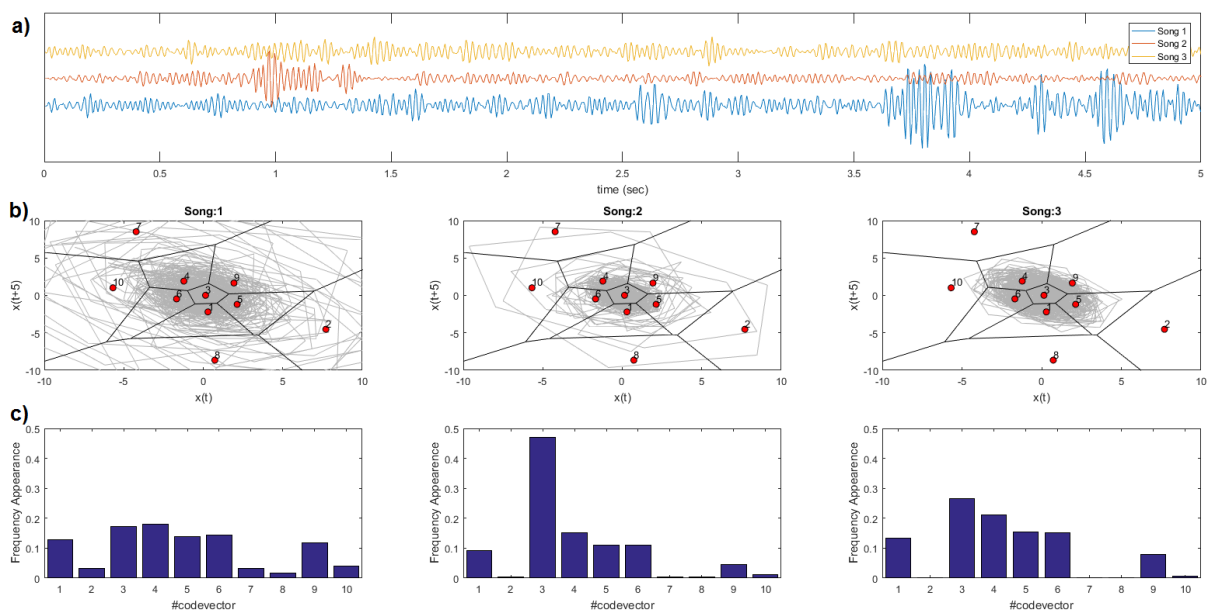

*Figure 2 A nonlinear dynamics approach to analyzing brain signal traces: a) music listening single-trial responses recorded at FP1 sensor filtered within $\gamma_{low}$ frequency band, b) embedding of the corresponding dynamics as (approximate) trajectories in a 2D space, with the Voronoi regions defined based on a given set of 10 reference points (code-vectors), denoted with red disks. c) Describing the trajectories as histograms over the given codebook (Voronoi regions are treated as bins).*

Given a training set of signals, $s_1(t), s_2(t), \ldots, s_M(t)$, we first form the corresponding trajectory matrices $\mathbf{Tr}\{s_1(t);d_e,\tau\}$, $\mathbf{Tr}\{s_2(t);d_e,\tau\}$,…, $\mathbf{Tr}\{s_M(t);d_e,\tau\}$, and then create an overall data-matrix, by stacking them : $\mathbf{X^{data}}=[\,\mathbf{Tr}\{s_1(t);d_e,\tau\}\,|\,\mathbf{Tr}\{s_2(t);d_e,\tau\}\,|\ldots|\,\mathbf{Tr}\{s_M(t);d_e,\tau\}\,]$. The $[M \cdot N \times d_e]$ data matrix is then fed to Neural Gas, a self-organized network that efficiently converges to a set of $k \ll M \cdot N$ codebook vectors using a stochastic gradient descent procedure with a ''soft-max'' adaptation rule that minimizes the average distortion error. The derived codevectors $\mathbf{c_j} \in \mathbb{R}^{d_e}$, j=1,2,…k constitute the initial set of reference brainwave patterns for the encoding of single trial brain dynamics. In principle, we use a relative high value for the codebook size, e.g. k=100 or 200. Such a value is needed to ensure that the reconstructed space will be partitioned at a sufficient detail. This in turn will enable the

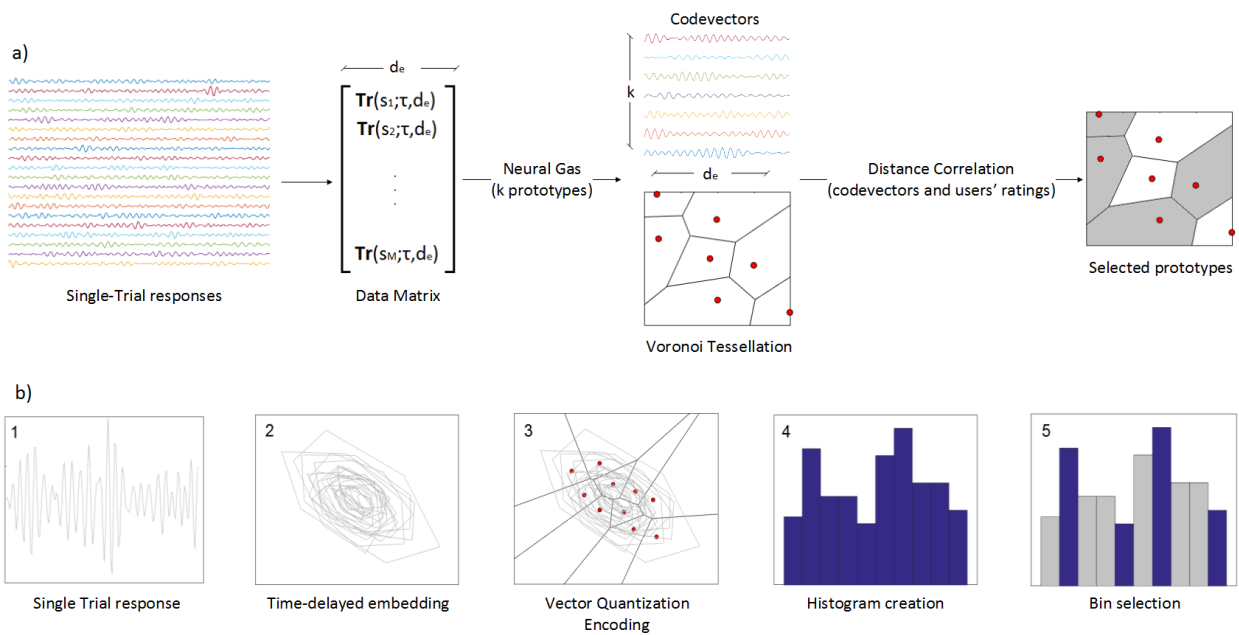

*Figure 3* a) Designing and editing the codebook based on training data. b) The steps of VQ encoding of a single-trial response. 1-2: the band-pass filtered trace undergoes time-delayed embedding; 3: Phase-space is divided into regions via Voronoi tessellation and all vectors are assigned to their nearest prototype. 4: The encoded trajectory is represented in histogram format. 5: The 'projections' within the most informative regions are kept for the Biomarker.

accurate representation of the dynamics after the encoding and will provide sufficient flexibility in the quest for informative Voronoi regions.

Next, the VQ-encoding is applied to the training set of signals. In this way, a signal $s_i(t)$ is transformed to a distribution pattern $\mathbf{h}_i=[h_i(1), h_i(2)…,h_i(k)]$, with the $r^{th}$-entry reflecting the relative activation of the $r^{th}$ Voronoi region during listening to the musical piece associated with the $i^{th}$ signal. The associated labels $score_i$, $i=1,2…,M$ expressing the subjective evaluation of the listener for each song, are utilized in a feature selection procedure that follows the principles of dynamic programming and exploits the generic character of the Distance Correlation measure (described in section 3.2). After the initial ordering of the individual codevectors based on their dependence with the level of music appreciation, i.e. based on the quantity $R(h_i(r), score_i)$, an incremental procedure is followed. It starts with the codevector $r_{[1]}$, the measurements $h_i(r_{[1]})$ of which show the highest nonlinear dependence. At each iteration, the ranked list of codevectors is traversed systematically and the current list of selected codevectors is augmented with the next in order codevector only if the overall measure of dependence increases. The procedure terminates when no further increase is achieved and returns a selected list of codevectors {r[1], r[2],... r[sel]}. The selected codevectors, named hereafter as *codewaves*, constitute the basis for the new biomarker. Fig.4 exemplifies the step of selecting the codewaves by means of a semantic map representation [25], (a 2D scatter plot derived based on multidimensional scaling) for signals recorded at FP1 and band-pass filtering within γ band.

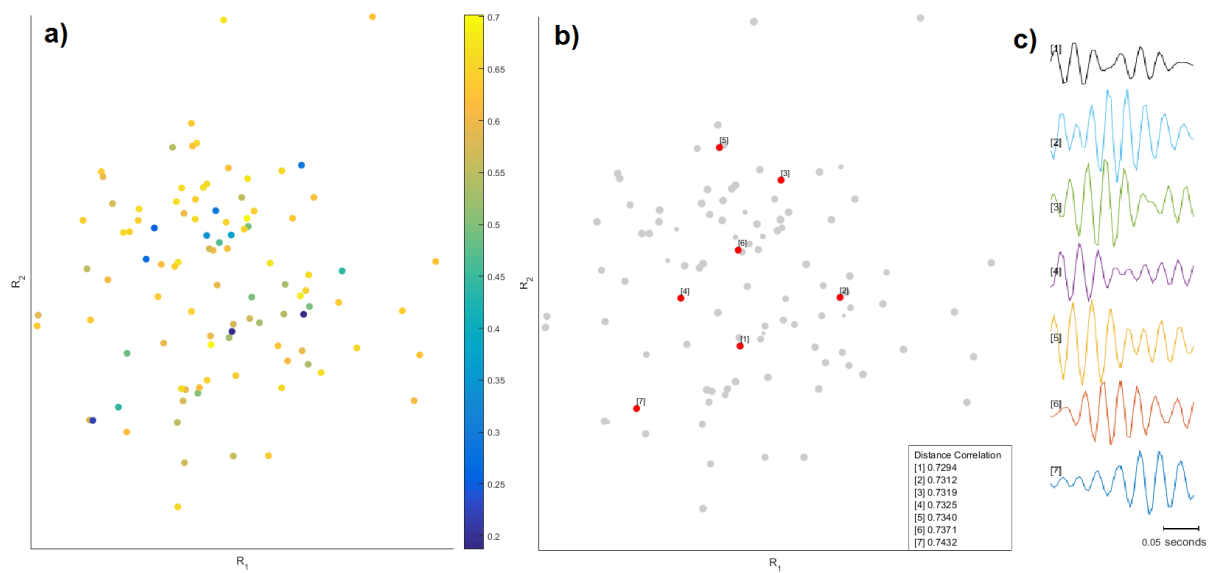

*Figure 4* Selecting CodeWaves: a) a 2D embedding of the initial codebook which consists of 100 code-vectors coloured by the distance correlation (with the subjective score of music appraisal) of each one separately, b) the spatial representation of the codebook editing step with the selected codewaves ranked by their distance correlation value (legend refers to the score of the hierarhically formed list) c) the seleted codewaves.

Wrapping up section 3.2, the nonlinear dynamics biomarker (after the Codebook design and editing) is implemented with the steps depicted graphically in Fig.3b.

## 3.3 Distance Correlation

In statistics and in probability theory, Distance Correlation is a measure of statistical dependence between two random variables or two random vectors of arbitrary, not necessarily equal, dimension. Distance Correlation, denoted by R, generalizes the idea of correlation and holds the important property that R(**x**,**y**)=0 if and only if **x** and **y** are independent. R index satisfies 0≤ R ≤1 and, contrary to Pearson's correlation coefficient, is suitable for revealing non-linear relationships [26]. In this work, it was the core mechanism for identifying neural correlates of subjective music evaluation, by detecting associations between the results of signal descriptors (or combinations of them) and the listener's scores.

## 3.4 Extreme Learning Machines

Machine learning deals with the development and implementation of algorithms that can build models able to generalize a specific function from a set of given examples. Regression is a supervised learning task that machine learning can handle with efficiency of similar, or even higher, level than the standard statistical techniques, and -mainly- without imposing hypotheses. In this work, the decoding of subjective music evaluation was cast as a (nonlinear) regression problem. A model was then sought (i.e. learned from the experimental data) that would perform the mapping of patterns derived from the brain activity descriptors to the subjective music evaluation.

ELMs appeared as a suitable choice due to their documented ability to handle efficiently difficult tasks without demanding extensive training sessions [27]. They are feedforward ANNs with a single layer of hidden nodes, where the weights connecting inputs to hidden nodes are randomly assigned and never updated [28].

# 4 Implementation and Experiments

Our experimentations evolved in two different directions. First, a set of experiments was run so as to use the experimental data for establishing the brainwave pattern(s) that would best reflect the music appraisal of an individual and train the learning machine from music pieces of known subjective evaluation. Next, additional experiments were run that implemented the real-time scoring by the trained ELM-machine so as to justify the proposed BCIs in a more naturalistic setting. In both cases, the musical pieces were delivered through a popular music on-demand streaming service (Spotify) that facilitates the registration of the listeners' feedback ('like'-'dislike') to adapt the musical content to their taste and make suggestions about new titles.

## 4.1 Participants

All 5 participants were healthy students of AUTH Informatics department. Their average age was 23 years and music listening was among their daily habits. They signed an informed consent after the experimental procedures had been explained to them.

## 4.2 Data Acquisition

Having in mind the user-friendliness of the proposed scheme, we adopted a modern commercial dry-sensor wireless device (i.e. Interaxon's Muse device - http://www.choosemuse.com) in our implementations. This "gadget" offers a 4-channel EEG signal, with a topological arrangement that can be seen in Fig.5. The signals are digitized at the sampling frequency of 220 Hz. Data are transmitted under *open sound control* (OSC) protocol, which is based on UDP/IP to facilitate communication and flexible interoperability among computers, sound synthesizers and other multimedia devices optimized for modern networking technology [29].

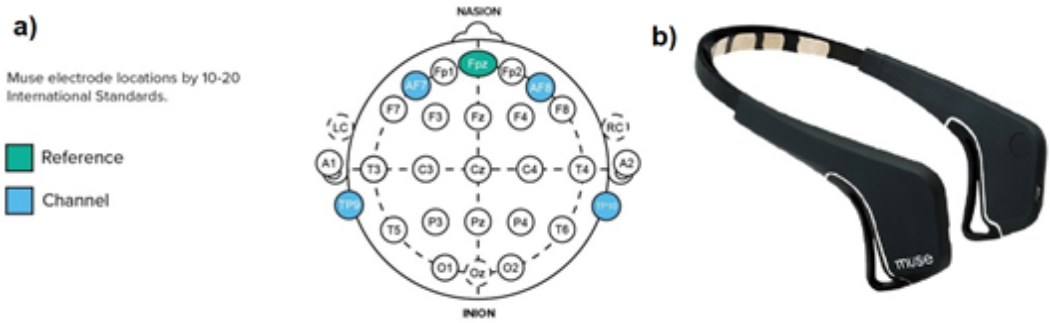

*Figure 5 a) Topological arrangement of available Electrodes b) Interaxon's Muse headset.*

## 4.3   Experimental Procedure

Prior to placing the headset, subjects sat in a comfortable armchair and volume of speakers was set to a desirable level. They were advised to refrain from body and head movements and enjoy the music experience. Recording was divided into sessions of 30 minutes duration. The music streaming service was operated in radio mode, hence randomly selected songs (from the genre of their preference) were delivered to each participant while his/her brain activity was registered. Among the songs there were advertisements. That part of recordings was isolated from the rest. The recording procedure was integrated, in MATLAB, together with all necessary information from the streaming audio signal (i.e. song id, time stamps for the beginning and termination of each song). Participants evaluated each of the listened song, using as score one of the integers {1,2,3,4,5}, during a separate session just after the end of the recording. These scores, together with the associate patterns extracted from the recorded brain signals, comprised our training set. The overall procedure was repeated on a different day. On average, the recorded activity was corresponding to 30 songs per participant and approximately 40% were unknown to the listeners. In addition to the above data, two of the subjects participated in an extra recording session, during which the trained BCI-systems operated in real-time and their predictions were compared with the scores provided by the listeners just after the end of the recording.

## 4.4  Data Analysis

The preprocessing of signal included 50Hz component removal (by a built-in notch filter in *MuseIO* - the software that connects to and streams data from Muse), DC offset removal, and removal of the signal segments that corresponded to the 5 first second of each song (in order to avoid artificial transients).

The digital processing included band-pass filtering for deriving the brainwaves of standard brain rhythms and computation of the descriptors described in sections 3.1 and 3.2. To increase frequency resolution, we divided the β rhythm into $β_{low}$ (13-20 Hz), and $β_{high}$ (20-30 Hz) sub-bands and derived descriptors separately. Similarly, the γ rhythm was divided into $γ_{low}$ (30-49Hz) and $γ_{high}$ (51-90Hz). As a means to increase the volume of available data, multiple segments were extracted from the bandpass-filtered brain activity corresponding to listening to a single song listening. These extracts had been selected randomly, with their length parametrized during the off-line experimentation between 10 and 100 seconds. This 'bootstrapping' step was compatible with the possibility that, in radio-listening mode, the listening may start in the middle of a song.

The two types of descriptors (presented in 3.1 and 3.2 respectively), were applied to the available single-trial segments, in order to identify the optimal combination (or subset in the latter case) of features to be included in the proposed BCI-systems. In the case of NeuroPicks, the feature selection step was performed collectively for all participants, resulting in a single composite Biomarker. On the contrary, such a procedure was not compatible with the nature of NeuroPicks$^{VQ}$ that relies on the adaptive design of codewaves. Hence, the nonlinear dynamics biomarker was defined on a subject-by-subject basis.

The derived biomarkers were utilized as ''feature-extractors'' and a training set of patterns was gathered by estimating the biomarkers from additional (independently sampled) extracts of brain activity. These patterns were employed for tailoring an ELM model to each participant. This was considered the "overall training" phase for the supervised component(s) of the BCI systems. In an ''in-situ'' testing phase (section 5.3), the subject-specific ELM model was applied to streaming data (from the biomarkers operating on line) in order to predict the listener's evaluation.

# 5 Results

## 5.1 NeuroPicks : the composite biomarker + ELM

The biomarker in this BCI-system was designed based on a feature-engineering procedure that was meant to be universally valid, i.e. to result to a biomarker that would be applicable to any user, without the need for time-consuming adaptations. Towards this end, the ensemble of descriptors described in 3.1 was derived regarding all defined brain rhythms (7), available sensors (4), both pairs of symmetrically placed sensors (4), and legitimate pairs of cross-frequency interactions $rhythm_{low} \rightarrow rhythm_{high}$ (21). All these measurements, from the single-trial segments with the highest length, were gathered in a *tall* matrix with #rows=(30songs x 10extracts x 5subjects) and #columns=140. The statistical dependence of each one dimension (column) with the associated scores of subjective music evaluation was estimated based on the Distance Correlation measure and a bootstrapping scheme during which 100 random samples (from the rows) were created and the individual estimates were averaged. Fig.6, includes the obtained correlation measurements in the form of three distinct tables, split according to the type of each information dimension. The 140 features were ranked in descending order according to R-score (i.e. regarding their music evaluation expressiveness) and a dynamic programming methodology was then applied. Starting with the feature of highest R, we traversed systematically the ranked list for the combination that would eventually maximize the Distance Correlation. This procedure led to a set of 5 features (integrated over time) that constituted the synthesized music appraisal biomarker. This biomarker included the normalized temporal asymmetry index in $β_{low}$ band, relative energy of α band at temporal electrodes, $γ_{low} \rightarrow γ_{high}$ PAC at FP1 sensor and relative energy of θ band at TP9.

The relevance of the designed biomarker to music evaluation showed a dependence on the length of segment based on which it had been evaluated. Figure 7 indicates this trend (as averaged profile across participants). It can be observed that while the performance of the biomarker is poor when fed with short EEG segments of music listening, its effectiveness constantly increases with the duration of the listened music. This trend provided an initial indication about how fast the automated music evaluation system NeuroPicks could operate.

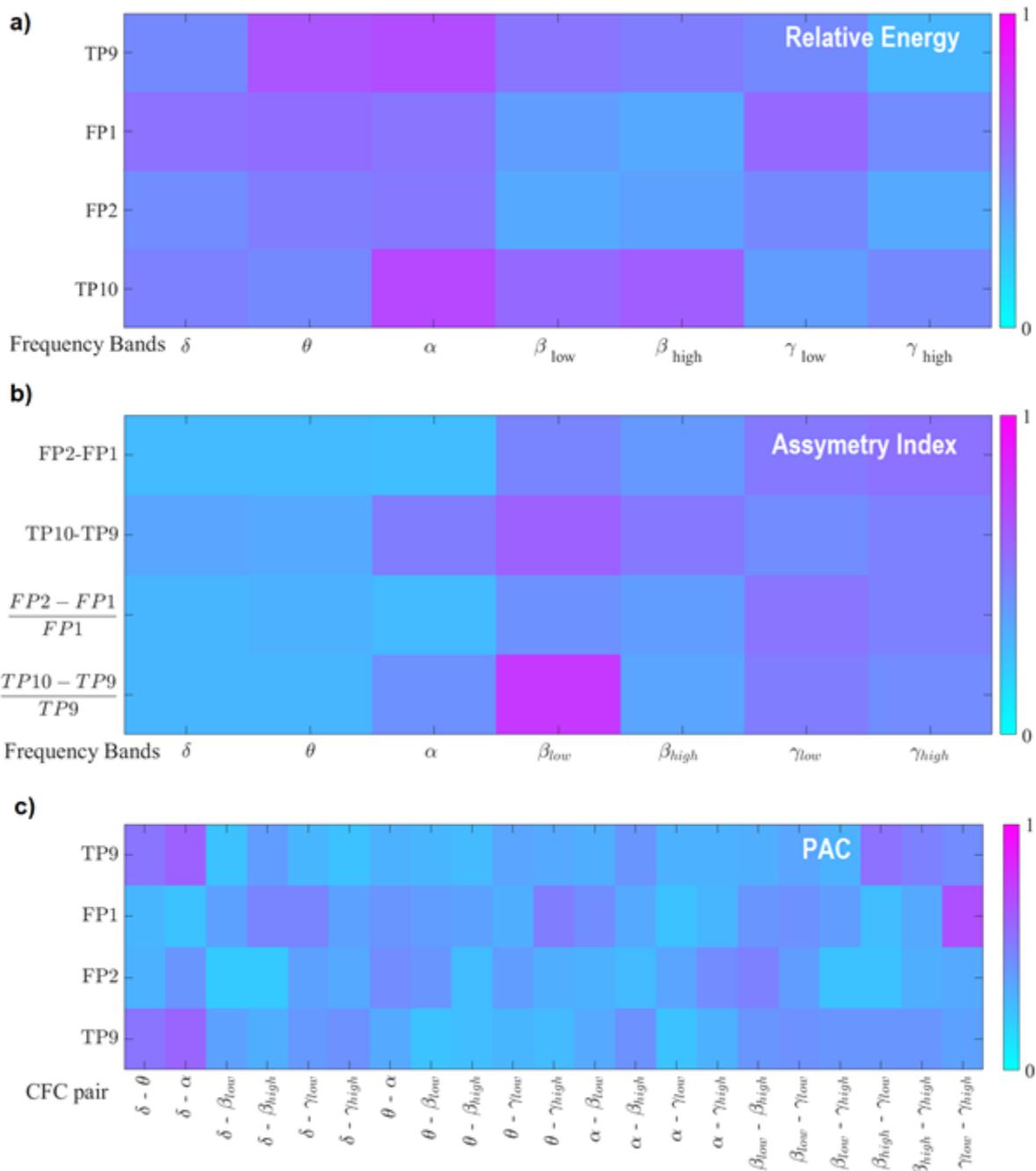

*Figure 6* *Distance Correlation score of all signal descriptors, derived based on 90-sec long segments of brain activity during music listening.*

During the offline experimentation, the available data (the biomarker patterns of each of the five participants along with the associated ratings) were randomly partitioned in training and testing set, in a 60%-40% proportion for a Monte-Carlo cross validation. An ELM was trained using the former set and its performance was quantified using the latter set. The number of neurons in the hidden layer, the only parameter to be tuned in ELMs, was selected as the smallest number of neurons such that the training and testing error were converging at an acceptable level, lower than 0.01. This number was found ranging from 12 to 18. The overall procedure was repeated 10 times, and averaged results were reported. The normalized root mean squared error (nRMSE), as defined for regression tasks, was found to be 0.063±0.0093 (mean±std across participants).

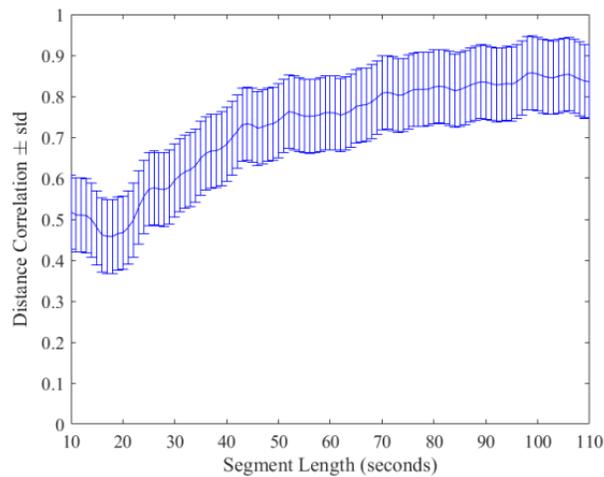

*Figure 7* *Distance Correlation of the composite biomarker with the score of subjective music appreciation as a function of the segment length.*

For comparison purposes, we also employed Support Vector Machines (SVMs), which performed slightly inferiorly. Although the difference was marginal, the very short training time was another factor in favor of employing ELMs.

## 5.2   NeuroPicks$^{VQ}$:  the adaptive biomarker + ELM

This music evaluation BCI system stemmed from the need to provide an indication about the listener's appraisal within few seconds, i.e. without postponing estimation till the end of the song, which is the ideal scenario for the previous system. It operates on short segments of brain activity and owns its efficiency to the adaptive character of the employed descriptor.

The most important operational difference (with respect to NeuroPicks), is that before the realization of NeuroPicks[VQ], training is required for the biomarker as well. Data learning is involved for optimizing the representation of brainwaves in a subject-specific manner. As a means to confine the search space, we examined the suitability of each sensor and brain rhythm for leading to a VQ-based biomarker that reflects well the subjective music evaluation, and parametrized this search over the length of the extracted segments (aiming for the shortest segments that could support acceptable performance). In every case, for the initial step of forming trajectories via time-delay embedding, the following strategy was followed for defining the two embedding parameters (deviating from the standard practice in the literature of nonlinear dynamics [30]). Having in mind to provide a detailed description, we always kept time delay τ=1 and leave the embedding dimension $d_e$ as an additional parameter to be optimized across subjects. For the step of initial codebook design, k was set to 100 (and experimentally verified that it was not a critical parameter as long as it remained above 70). The codebook-editing step usually led to the selection of 8 (range: (6-12)) codewaves depending on the situation (subject, sensor, rhythm and initialization of Neural-Gas algorithm). The statistical dependence of the designed biomarkers with the associated scores of subjective music evaluation was first estimated for each subject independently, based on the Distance Correlation measure and a bootstrapping scheme, and the individual results were then averaged across subjects. Fig.8, represents the suitability of each sensor and brain rhythm to provide sufficient information based on 20-sec segments and when the embedding dimension was $d_e$ =50. It can be observed that it was the activity in $\gamma_{low}$ from FP1 electrode that would lead to the most suitable biomarker. The corresponding correlation level with the subjective score was reach the value of 0.72. It should be mentioned here that the use of longer segments did lead to higher R-values, but the increase of R-values with time interval was building very slow and never exceed the value of 0.78.

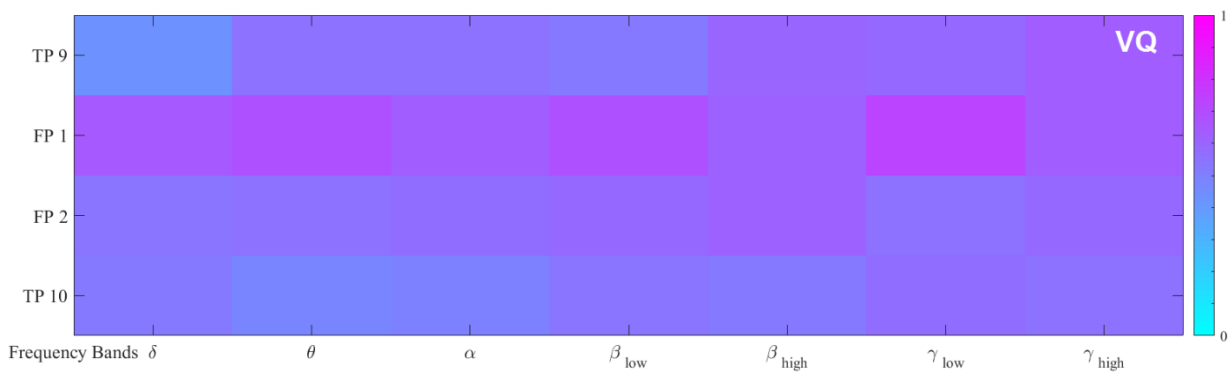

*Figure 8* Distance Correlation of the VQ-based biomarker (with the subjective music appreciation) for all electrodes and brain rhythms based on 20sec-long segments of brain activity signals during music listening. The shown measurements are the averaged across subjects Distance Correlation values.

Considering the above parameters as the prescribed values (i.e. sensor:FP1, rhythm: $\gamma_{low}$, segment:20sec, $d_e$=50), and working independently for each subject, we derived the VQ-biomarker for all (approximately 300=30 songs × 10 extracts ) available segments. The included codewaves had been defined in a personalized fashion and even their number was differing among subjects. The VQ-biomarker patterns were randomly partitioned in training and testing set, in a 60%-40% proportion for a Monte-Carlo cross validation. An ELM was trained using the former set and its performance was quantified using the latter set. The number of neurons in the hidden layer, was selected as described in 5.1. The overall procedure was repeated 10 times, and the normalized RMSE, averaged across repetitions, provided an indication of performance. By averaging across participants, the error for the NeuroPicks$^{VQ}$ system was estimated at 0.115 ±0.046. This is a satisfactory figure of performance considering the 20sec-interval needed.

## 5.3 Online evaluation of the BCI - systems

Additional experiments were performed in order to test the performance of the two systems in a realistic setting. Two of our subjects participated in additional recording sessions during which the (already custom-made) NeuroPicks systems were providing a read-out of the subjective music evaluation. In the case of NeuroPicks /NeuroPicks$^{VQ}$ , 90/20 seconds long segments were used. The main purpose of this setup was to truly evaluate the performance of the proposed BCIs on new, previously unseen data. Based on three recording sessions lasting for 90 minutes in total, 18 and 21 songs were delivered for the first and second subject, through the Spotify platform. The automated evaluation scores (predictions of subjective appraisal) were registered and compared with the listeners' ratings provided at the end of each recording session. The normalized RMSE for the NeuroPicks system was estimated at 0.09 and 0.07 for the two individuals. These estimates were based on the score predicted by the system using the activity during the initial 90 seconds of each song. The same performance index was estimated at 0.19 and 0.16 for the NeuroPicks$^{VQ}$ system. These estimates were based on a single score (mean value) produced by integrating along time the time-course of predictions produced by the system (see Fig.9).

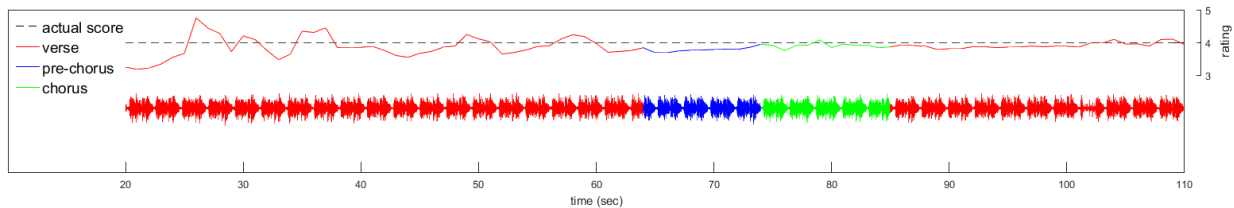

*Figure 9* Time evolving Predictions of NeuroPicks$^{VQ}$ BCI music evaluation system acting on recorded brain activity during listening to a song with high subjective rating. The corresponding audio signal is shown beneath, segmented into verse, pre-chorus and chorus.

# 6   Discussion

The paper reports our attempts to associate the listener's brainwaves with the subjective aesthetic pleasure induced by music. It also serves as a proof-of-concept study on the feasibility of interfacing a modern consumer EEG device with a popular on-demand music streaming service (Spotify).

Our results indicate that descriptors of the signals recorded from a restricted number of sensors (located over frontal and temporal brain areas) can be combined in computable biomarkers operating on long or short temporal segments and reflecting the listener's subjective music evaluation. The derived biomarkers' patterns can be efficiently decoded by regression-ELM leading to reliable readouts directly from the listener. The main advantage of the approach is that it complies with idea of employing EEG-wearables in daily activities and is readily embedded within the contemporary on-demand music streaming services.

Nevertheless, the problem of artifacts (noisy signals of biological origin) has not been addressed yet. For the presented results, the participants had been asked to limit body/head movements and facial expressions and as much as possible. Hence, before employing such a system to naturalistic recordings, methodologies for real-time artifact suppression (as in [31]) have to be incorporated.

Today, we live in the world of Internet of Things (IoT) where the interconnectedness among devices has already been anticipated and supportive technologies are now being realized [32]. However, to achieve a seamless integration of technology in people's lives, there is still much room for improvement. In a world beyond IoT, technology would proactively facilitate people's expectations and wearable devices would transparently interface with systems and services. To enable such scenarios would require to move from the IoT to the Internet of People (IoP) [33]. People would then participate as first-class citizens relishing the benefits of holistic human-friendly applications. As such, we present a pragmatic use of a consumer BCI that ideally fits in the anticipated forms of the digital music universe and demonstrate a first series of encouraging experimental results.

## Acknowledgements

This research did not receive any specific grant from funding agencies in the public, commercial, or not-for-profit sectors.